    \documentstyle[prl,aps,multicol,epsf]{revtex}

\begin{document}


\title{The Construction of a Quantum Markov Partition}

\author{Ra\'ul O. Vallejos$^1$ and Marcos Saraceno$^2$}

\address{{$^1$}Instituto de F\'{\i}sica, 
   Universidade do Estado do Rio de Janeiro, \\
   Rua S\~ao Francisco Xavier 524, CEP 20559-900 Rio de Janeiro, 
   Brazil \\
   {\rm e-mail: raul@dfnae1.fis.uerj.br} \\
   $^2$Departamento de F\'{\i}sica, Comisi\'on Nacional de
   Energ\'{\i}a At\'omica, \\
   Avenida del Libertador 8250, 1429 Buenos Aires, Argentina \\
   {\rm e-mail: saraceno@cnea.tandar.gov.ar}}

\date{\today}

\maketitle


\begin{abstract} 
We present a method for constructing a quantum Markov partition.  Its
elements are obtained by quantizing the characteristic function of the
classical rectangles.  The result is a set of quantum operators which
behave asymptotically as projectors over the classical rectangles
except from edge and corner effects.  We investigate their spectral
properties and different methods of construction.  The quantum
partition is shown to induce a symbolic decomposition of the quantum
evolution operator.  In particular, an exact expression for the traces
of the propagator is obtained having the same structure as Gutzwiller
periodic orbit sum.
\end{abstract}

\draft

\begin{multicols}{2}

\narrowtext


\section{Introduction}


For classical hyperbolic systems, symbolic dynamics provides
the proper coordinates for an efficient description of the chaotic
behavior \cite{devaney89}. Such description does not exist at the
quantum level (with the exception of a few important semiclassical
treatments \cite{bogomolny92}). This work is an attempt to apply
the techniques of symbolic dynamics in quantum mechanics. The ultimate
goal of this kind of investigations is to rewrite the equations of
quantum mechanics in terms of adequate symbols for a given (chaotic)
problem.

Symbolic dynamics requires a partition of phase space in various
regions.  We are thus faced with the problem of defining properly the
quantum analogues to bounded regions of phase space. The essential
difficulties for doing this are the limitations imposed by the
uncertainty principle.  Strictly speaking, quantum mechanics is not
only in contradiction with the notion of a phase space point but also
with that of a finite subset of phase space.

In a previous paper \cite{saraceno94} a symbolic decomposition along
these lines was studied, but no special constructions were necessary
because the invariant manifolds were aligned with the coordinate axes,
thus turning the elements of the generating partition into simple
projectors. Here we generalize the method of \cite{saraceno94} by
constructing certain objects (we call them {\em quantum rectangles})
which are the quantum equivalents to the classical elements of a
generating partition. Then we investigate their properties and
different possibilities for their construction.  The quantum rectangles
behave approximately as projectors over the corresponding classical
regions except from diffraction effects which are characteristic of
quantum phenomena.

Once the quantum rectangles have been defined, it is straightforward to
construct a quantum generating partition. In perfect analogy with the
classical case, this partition leads to a symbolic decomposition of the
propagator. Eventually, we obtain an {\em exact} trace formula having the
same structure as Gutzwiller's.

The rest of the paper is structured as follows. In Section II we argue
that the quantum analogue of a finite region of phase space can be
constructed in a natural way by simply quantizing the characteristic
function of that region. In Section III we show that in the
semiclassical limit the quantized regions display properties consistent
with the classical ones. Section IV describes the application of the
quantum generating partition to decompose the propagator. 
Finally, Section V contains the concluding remarks.


\section{Construction}


\label{section2}

The first step towards the construction of a quantum Markov partition
consists in defining the quantum analogue for a finite region $R$ of the
classical phase space (to be considered later as belonging to a
generating partition).  For the sake of simplicity, we restrict our
analysis to two dimensional phase spaces with the topology of a torus
(we further assume that the torus has unit area).
Extensions to spaces of higher dimensionality or to other topologies can
also be considered.  We want to construct an operator which is the
quantization of the characteristic function $\Delta_R$ of the region
$R$, 
\begin{equation} 
\Delta_{R}(q,p) = 
\left\{ 
\begin{array}{ll}
   1 & \mbox{if $(q,p) \, \epsilon \, R $}  \\ 
   0 & \mbox{otherwise}
\end{array} 
\right.  \; .
\end{equation} 
Let us just mention two simple properties of the characteristic
functions: distributivity with respect to the set intersection and
normalization
\begin{eqnarray}
\Delta_{R_1}\Delta_{R_2}   & = & \Delta_{R_1 \cap R_2} \; ,\\ 
\int dp \, dq \, \Delta_{R}(p,q) & = & {\cal A}_R      \; ;
\end{eqnarray}
the integral is over the torus and ${\cal A}_R$ is the area (volume) of
the region $R$.  For the moment these regions are arbitrary but
eventually they will become the elements of a partition of the phase
space.

To establish the connection with quantum mechanics we make use of a
phase space representation, that is, a basis
$\{\widehat{B}(q_k,p_j),\;1\le k,j \le N\}$ for operators acting on the
Hilbert space $\cal H$ of dimension $N=1/2\pi\hbar$ (the $q$ and $p$
representations on the torus are discrete, and mutually related through
a discrete Fourier transform \cite{saraceno90}).  Any operator
$\widehat{O}$ can be written as a linear combination of the elements of
the basis
\begin{equation} \label{basis}
\widehat{O}= \sum_{k,j=1}^N O(q_k,p_j) \, \widehat{B}(q_k,p_j)  ~.
\end{equation}
Conversely, for a given symbol $O(q_k,p_j)$, Eq.~(\ref{basis}) defines 
an operator $\widehat{O}$. We require the operator basis to decompose 
the identity
\begin{equation}
\sum_{k,j=1}^N \widehat{B}(q_k,p_j) = \openone_{\cal H} .~
\end{equation}
Two examples of operator bases will be considered: The Kirkwood
representation, associated to the basis 
$ \{ |q_k \rangle \langle q_k | p_j \rangle \langle p_j| \}$,
and a representation of projectors over coherent states,
$ \{ |q_k+ip_j \rangle \langle q_k+ip_j| \}$. 
In both cases the discretization used is $q_k=k/N$ and $p_j=j/N$, $1
\le k,j \le N$, corresponding to periodic boundary conditions on the
torus.  We construct the coherent set starting from a circular Gaussian
packet centered at $(1/2,1/2)$, say in the $q$ representation. Then,
this function is evaluated in the discrete $q$ mesh and normalized. The
whole set of coherent states is obtained by successive translations of
the initial state to all the points $(q_k,p_j)$ of the mesh
\cite{saraceno90}.

Both representations allow a natural construction of the quantization 
$\widehat{R}$ of a phase space region $R$:
\begin{eqnarray}
  \widehat{R}_K &=& \frac{1}{N}
                    \sum_{k,j=1}^N \Delta_R(q_k,p_j) 
            |q_k \rangle \langle q_k | p_j \rangle \langle p_j|  ~, \\
  \widehat{R}_z &=& \frac{1}{N^2}  
                    \sum_{k,j=1}^N \Delta_R(q_k,p_j) 
            |q_k+ip_j\rangle \langle q_k+ip_j| ~.
\end{eqnarray}
The normalization prefactors $1/N$ and $1/N^2$ are such that the
``quantum area" (to be defined later) of the whole torus is one. The
additional factor $1/N$ in the coherent case is due to the
overcompleteness of that representation.  While $\hat R_z$ is Hermitian
and treats symmetrically $p$'s and $q$'s, $\hat R_K$ is not. Therefore,
in applications we use the symmetrical combination
$\widehat{R}_K^s=(\widehat{R}_K + \widehat{R}^\dagger_K)/2$ 
(we come back to this point later).

By defining the operators as quantizations of the characteristic
functions of the classical regions we guarantee that they have the
expected semiclassical limit. We will show that the Gaussian rectangle
${R}_z$ tends smoothly to its classical counterpart. On the other side,
the convergence of ${R}_K$, which has been constructed from a sharp
distribution, shows characteristic rapid oscillatory structure.


\section{Properties}


\label{section3}

\begin{figure}[ht]
\hspace{-1.0pc} 
\vspace{ 1.0pc}
\epsfysize=7.5cm
\epsfbox[69 182 490 549]{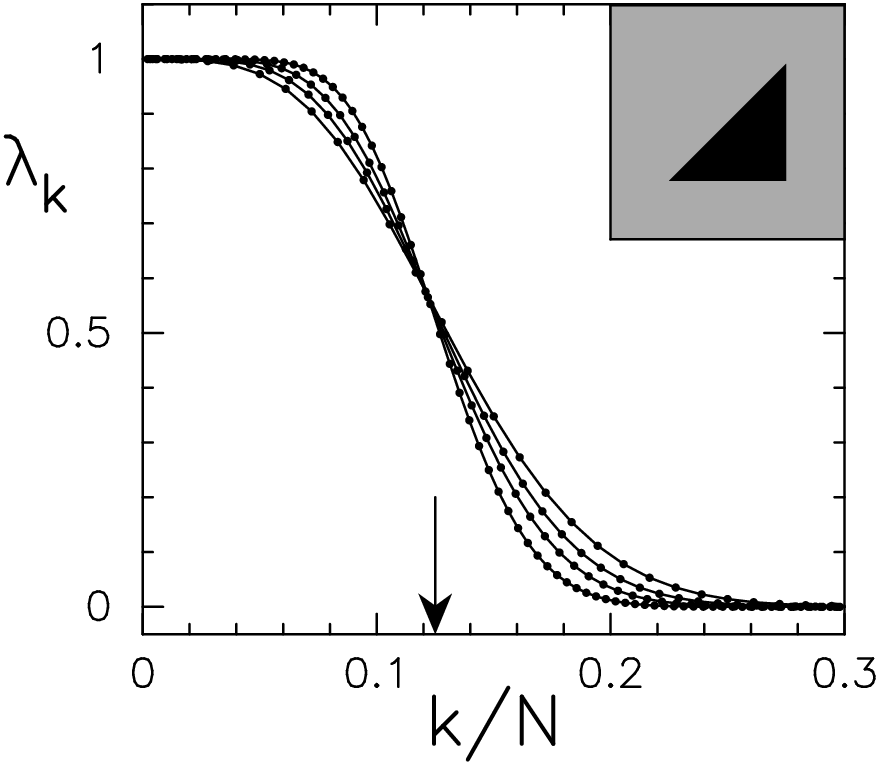}
\caption{Eigenvalues of a triangular region (inset) constructed from
the coherent basis. We plot the eigenvalues $\lambda_k$ 
as a function of the normalized eigenvalue number $k/N$, where
$N$=$1/2\pi\hbar$ is the semiclassical parameter. We have
considered $N$=90,120,160,240. 
Notice that as $N \to \infty$ the distribution of eigenvalues tends 
to a step function, the position of the step being associated with
the area of the classical region (indicated with an arrow).
The lines are a guide to the eye.}
\label{fig1}
\end{figure}

\begin{figure}[ht]
\hspace{-1.0pc}
\vspace{ 1.0pc}
\epsfysize=7.5cm
\epsfbox[94 182 459 554]{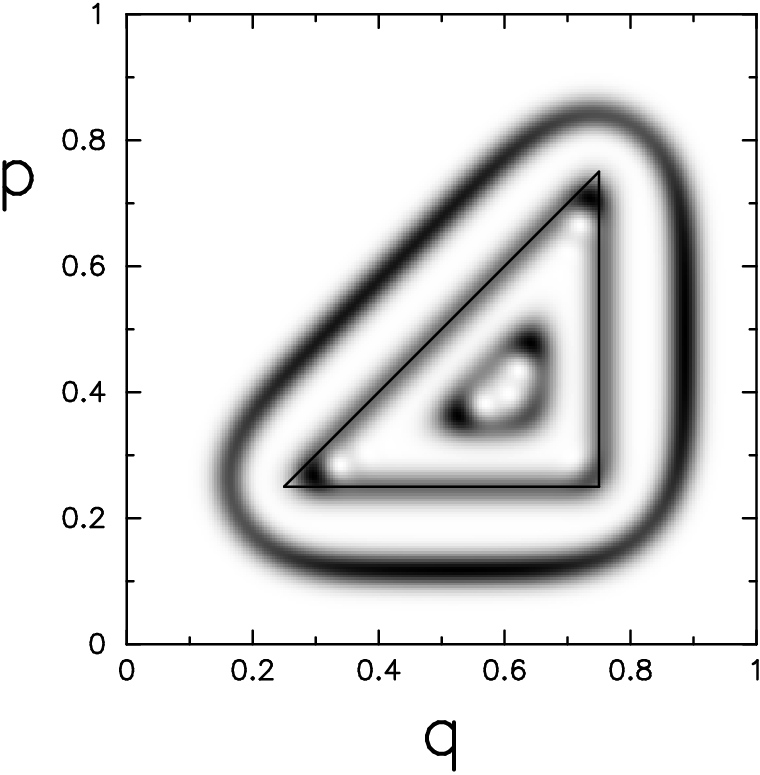}
\caption{Husimi representation of three eigenfunctions of the
triangular region of Fig.~\ref{fig1}. (This is a linear gray plot, 
with black and white corresponding to the highest and smallest 
amplitudes, respectively.) 
The associated eigenvalues
are $\lambda_k$=1.000,0.498,0.728; respectively $k$=4,31,90 ($N$=240).
The border of the classical region is shown for reference 
(full black line).}
\label{fig2}
\end{figure}

\begin{figure}[ht]
\hspace{-1.0pc}
\vspace{ 1.0pc}
\epsfysize=7.5cm
\epsfbox[69 190 480 549]{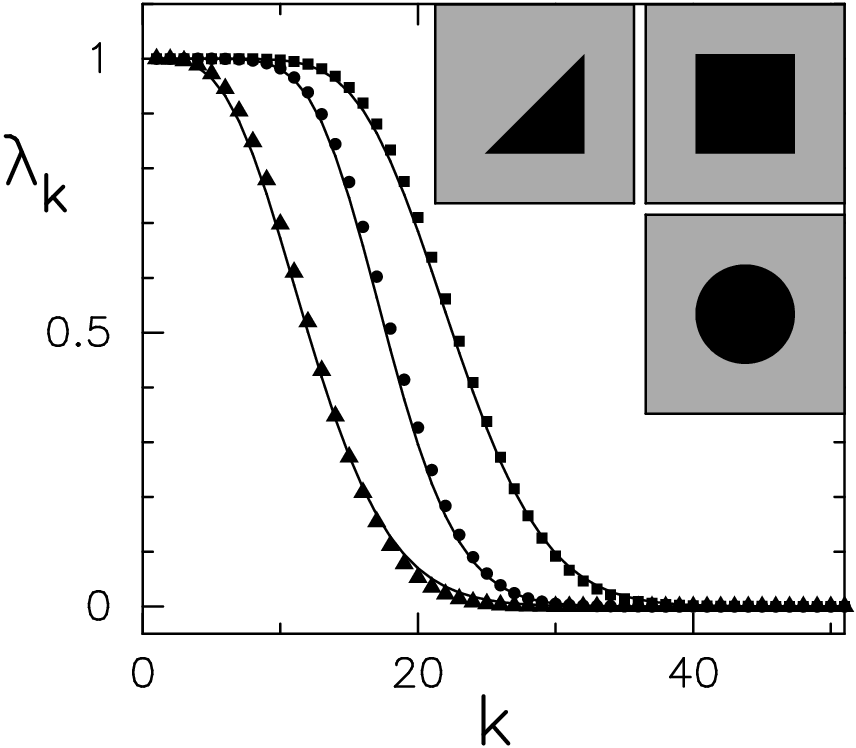} 
\caption{Eigenvalues of three regions: triangle, square, and circle (inset).
There is a one-to-one correspondence between symbols and regions.
The full lines are the analytical distributions discussed in the
text ($N$=90).}
\label{fig3}
\end{figure}

The spectral analysis of the quantum rectangles are the key to
understanding their general properties. We begin by studying the
Gaussian regions. For the case of a triangular region, Fig.~\ref{fig1} 
shows the way in which 
$\widehat{R}_z$ behaves in the limit $N \to \infty$. 
There we plot the eigenvalues $\lambda_k$ (associated to the eigenvectors
$|\psi_k \rangle$) in decreasing order.
Most of the eigenvalues take the values $\approx 0$ or $\approx 1$.  
Intermediate values exist, but their relative
number goes to zero in the semiclassical limit as a ratio surface to
volume.
Therefore, semiclassically, the rectangle behaves as a
projector. Figure~\ref{fig2} shows that the Husimi representations 
$|\langle q+ip| \psi \rangle|^2$ of the corresponding eigenfuctions 
are localized on nested triangles concentrically with the boundary of 
the classical region. 
The situation is very similar to that of integrable Hamiltonians, where
the Husimi density of an eigenfunction is localized over the associated
quantized torus and decays exponentially as one moves away from the
torus. Exploiting this analogy we can derive a semiclassical
quantization rule for the eigenvalues and eigenfunctions of a quantum
region.  Notice first that the eigenvalue equation for $\hat R_z$ is
\begin{equation}
\frac{1}{N^2}
\sum_{(q,p)} \Delta_R 
| q+ip \rangle \langle q+ip | \psi_k \rangle = 
                       \lambda_k | \psi_k \rangle    \; ,
\end{equation}
implying that  
\begin{equation}
\label{lambdak}
\frac{1}{N^2}
\sum_{(q,p) \in R} |\langle q+ip | \psi_k \rangle|^2 = \lambda_k \; 
\end{equation}
(the sum over the whole torus giving one).  Let's now make the
following assumptions.  Sums can be substituted by integrals (we are
interested in the limit $N \to \infty$).  The Husimi of the $k$-th
eigenfunction is associated to a quantized ``torus" lying at a distance
$d_k$ from the border of the region. The function $d_k$ depends on the
shape of the region and arises from packing $k$ quasi one-dimensional
strips of area $h$ concentrically with the border of the region,
starting from inside. Last, in the direction
perpendicular to the torus, $\hat y$ ($y=0$ on the torus), 
the Husimi is a normalized Gaussian:
\begin{equation}
\label{Gaussian}
                     \exp(-y^2/\hbar)/\sqrt{\pi \hbar} \; .  
\end{equation}
Combining Eqs.~(\ref{lambdak},\ref{Gaussian}) we arrive at the 
semiclassical quantization rule
\begin{equation}
\label{analytical}
\lambda_k = 
     \frac{1}{2} + 
     \frac{1}{2} \mbox{erf} \left( \frac{d_k}{\sqrt{\hbar}} \right) \; .  
\end{equation}
We give expressions for $d_k$ for the simplest-shaped regions:  
a square, the triangle of Fig.~\ref{fig1}, and a circle
\begin{equation}
d_k = \left\{
      \begin{array}{ll}
      \frac{L-\sqrt{kh}}{2}           & \mbox{square}   \\
      \frac{L-\sqrt{2kh}}{2+\sqrt{2}} & \mbox{triangle} \\
            R-\sqrt{\frac{kh}{\pi}}   & \mbox{circle}
      \end{array}
      \right. \; ,
\end{equation}
where $R$ is the radius of the circle and $L$ is the side of both 
the square and the triangle (see inset Fig.~\ref{fig3}).
 
In Fig.~\ref{fig3} we compare the analytical expression (\ref{analytical})
with the numerical results for three different regions, verifying that
the agreement is excellent, even for the relatively small $N=90$.

The quantization of classical regions with sharp boundaries by way of
coherent states presented above has the advantage of producing very
smooth and analytically understandable results. 
The sharp edges are blurred by the Gaussian smoothing and the
resulting quantum rectangles are always ``soft" on the scale of $\hbar$.

\begin{figure}[ht]
\hspace{-1.0pc}
\vspace{ 1.0pc}
\epsfysize=7.5cm
\epsfbox[69 165 490 549]{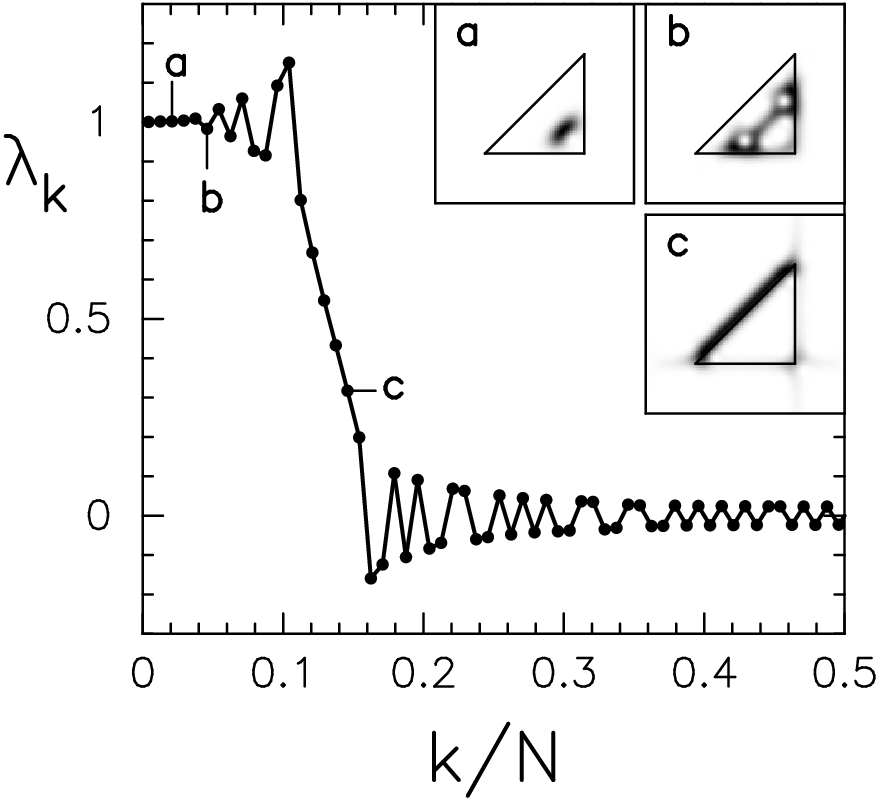} 
\caption{Eigenvalues of a triangular region quantized from the Kirkwood
basis, and three eigenfunctions 
(Husimi representation, linear gray plot). 
The ordering of the eigenvalues 
is such that the parameter $|1-|\lambda_k||$ increases to the right.
The eigenfunctions labelled {\bf a,b,c} correspond to the eigenvalues
$\lambda_k$=1.002,0.983,0.317 ($k$=3,6,18), respectively ($N$=120).}
\label{fig4}
\end{figure}

Other representations, namely Kirkwood and Wigner\cite{balazs84}, 
allow higher
definition but display characteristic diffraction effects at the edges
and corners.  Fig.~\ref{fig4} shows the eigenvalues of the operator
$\hat R_K^s$ of the triangular region.  Notice that the distribution of
eigenvalues is not smooth as in the coherent case but presents a
singularity associated to boundary effects. This singularity is
inherent to the sharpness of the Kirkwood construction and is also
displayed by the non-hermitian rectangles $\widehat{R}_K$ and
$\widehat{R}^\dagger_K$ (not shown). Some typical eigenfunctions are
also displayed (inset). In this case, the eigenfunctions do not present
the high degree of symmetry of the coherent case, but are rather
irregular.  The eigenvalue still determines the localization of the
eigenfunction with respect to the border ($<1/2$, interior; $>1/2$,
exterior).
However, as the eigenfunctions are not nested like in the coherent
case, the ordering is not always unambiguous.
 
Except for boundary effects, the Kirkwood rectangles behave 
asymptotically in the same way as the coherent ones, i.e., they tend to 
projectors over the classical regions.

Besides the nice spectral behavior discussed above, the quantum
rectangles (either coherent or Kirkwood) should display some additional
properties for our construction to be consistent: 

(a) How does one
define the ``area" of a quantum region? In order to quantify the
dissipation of a quantum Smale-horseshoe map, we argued in 
\cite{saraceno96}
that the usual operator norm 
$ \mbox{Tr} (\hat{R} \hat{R}^\dagger) $
is
a reasonable definition of area. For the Kirkwood rectangles $\hat R_K$ 
it is
easy to prove that this definition coincides exactly with the classical
area ${\cal A}_R$. Alternatively one could simply define area as
$\mbox{Tr} \hat{R}$, in which case classical and quantum
areas are identical for both representations. Anyway, as $\hat R$
tends to a projector
\begin{equation}
\mbox{Tr} (\hat{R} \hat{R}^\dagger) \approx  
\mbox{Tr}  \hat{R}  = {\cal A}_R \; .
\end{equation}
Thus both expressions are acceptable definitions 
of quantum area.

(b) For the study of spectral properties the Hermitian operator
$\widehat{R}^s_K$ was preferred to the non-Hermitian $\widehat{R}_K$
and $\widehat{R}_K^\dagger$. The latter are more appropriate for the
decomposition of the propagators we present in Section \ref{section4}.
However, in the limit ${\cal A}_R \gg \hbar$,
$\widehat{R}_K$ and $\widehat{R}^\dagger_K$ will be approximately
equal, given that they only differ in the ordering of $q$'s and $p$'s.
Then $\widehat{R}_K$, $\widehat{R}_K^\dagger$, and $\widehat{R}_K^s$
are semiclassically equivalent.

\begin{figure}[ht]
\hspace{-1.0pc}
\vspace{ 1.0pc}
\epsfysize=7.5cm
\epsfbox[94 182 465 555]{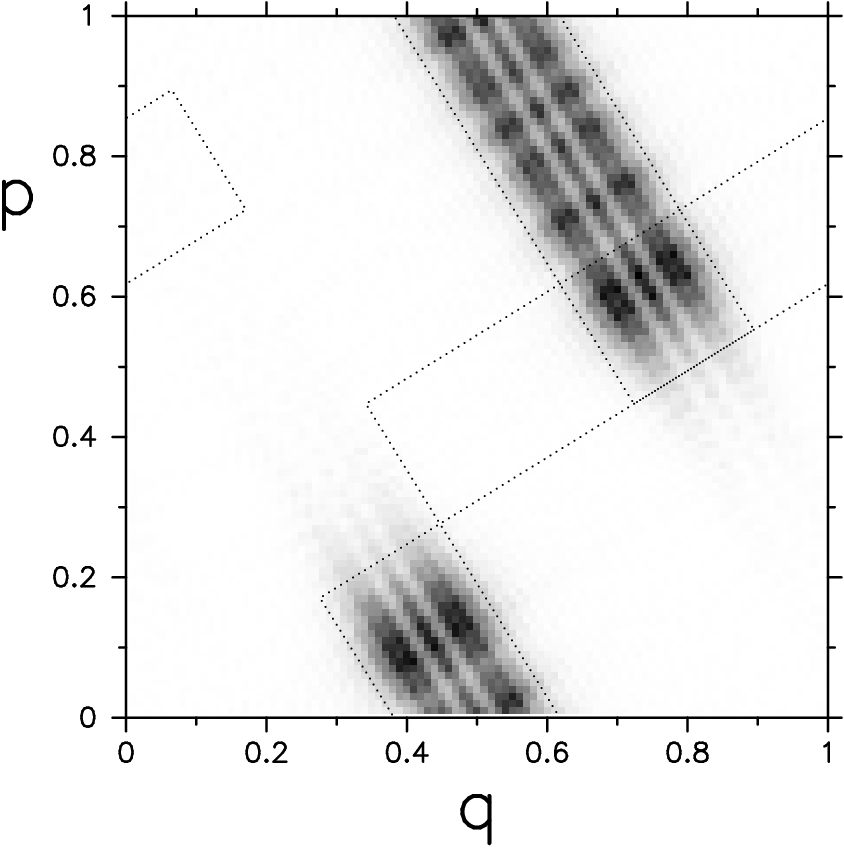}
\caption{Propagation of a Kirkwood rectangle. We show a linear
gray plot of
$|\langle p | \hat U^\dagger \hat R_1 \hat U | q \rangle |$, where
$R_1$ is one of the five elements of the generating partition
of Arnold's cat map (see Fig.~\ref{fig6}) 
and $\hat U$ is the quantized cat map (Section \ref{section4}).
Compare with the boundaries of the classically propagated 
rectangle, $U^{-1}(R_1)$ (dots). For the sake of future
referencing we also display the boundary of the element $R_5$ 
(dots). The dimension of the Hilbert space is $N$=100.}
\label{fig5}
\end{figure}

(c) Quantization and propagation must commute: 
If $U$ is a classical simplectic map and $\widehat{U}$ its 
quantization, then
\begin{equation}
  \widehat{U}^T \widehat{R} \widehat{U}^{-T} \to 
  \widehat{U^T(R)} \; 
\end{equation}
where it is understood that one must fix $T$ and take the limit 
$\hbar \to 0$. To illustrate the way in which this limit may 
be reached we show in Fig.~\ref{fig5} the propagation of a Kirkwood
element of the generating partition of Arnold's cat map 
(see Fig.~\ref{fig6}).
Notice that besides the bulk classical propagation, diffraction effects
associated to the edges and corners are clearly visible.
We remark that this behavior is typical of sharp representations.
Coherent rectangles behave in a much smoother way.

(d) We also expect quantization to commute with the classical set 
operations:
\begin{eqnarray}
 & \widehat{R_1 \cap R_2} \approx 
   \widehat{R}_1 \widehat{R}_2      \approx 
   \widehat{R}_2 \widehat{R}_1 \; ,& \\
 & \widehat{R_1 \cup R_2} \approx 
   \widehat{R}_1+\widehat{R}_2 - \widehat{R}_1 \widehat{R}_2 \; .
\end{eqnarray}
In the next section we show an application of the quantum regions 
which is an indirect test of the validity of these statements.

\section{Symbolic decomposition of the traces of the propagator}

\label{section4}

Before discussing the applications of the quantum rectangles in quantum 
dynamics, we present a short reminder of classical symbolic dynamics
in a setting appropriate to the transition to quantum mechanics.

For a hyperbolic map $U$, symbolic dynamics relates the orbits of $U$ to
symbolic sequences by means of a partition of phase space.
Such partition consists of a set of regions $R_1,R_2,\ldots,R_P$
(usually called ``rectangles'') which satisfy the following (Markov)
properties.
The boundaries of $R_i$ are defined by segments of the expanding and
contracting manifolds of $U$. Whenever $U(R_i)$ intersects the interior
of $R_j$, the image cuts completely across $R_j$ in the unstable direction.
Similarly, the backwards image $U^{-1}(R_i)$ cuts completely across
the other rectangles along the stable direction.
\cite{devaney89,cornfeld80}.

Once one has constructed the Markov partition, successively finer
partitions are obtained by intersecting the elements of the basic 
partition with its positive and negative images by the map 
(product partition):
\begin{equation}
R_{\epsilon_{-K} \ldots \epsilon_{-1} \cdot \epsilon_{0} 
          \epsilon_{1} \ldots \epsilon_{M}}
          =\bigcap_{s=-K}^{s=M} U^{s} (R_{\epsilon_s}) \; , 
\label{product}
\end{equation}
where $\epsilon_s$ can take any of the values $1,2,\ldots P$. 
Each element of the new partition can be labeled by a different symbolic 
code 
\begin{equation}
\nu(-K,M)=
\epsilon_{-K} \ldots \epsilon_{-1} \cdot
                     \epsilon_{0} 
                     \epsilon_{1} 
              \ldots \epsilon_{M} \; .
\label{code}
\end{equation}
As the original rectangles, the rectangles above possess the property 
of decomposing the phase space into disjoint regions (we do not take into
account borders, which are zero-measure).  
When acting on these rectangles, the map is simply a {\em shift\/}:
\begin{equation}
U^{-1}
(R_{\epsilon_{-K} \ldots \epsilon_{-1} \cdot 
    \epsilon_{ 0} \epsilon_{1}\ldots \epsilon_M })=
 R_{\epsilon_{-K} \ldots \epsilon_{ 0} \cdot 
     \epsilon_{1} \ldots \epsilon_{ M} } ~.
\end{equation}
If, in the limit $K,M \to \infty$, each element of the product
partition is a single point, the code is said to be 
{\em complete}. It may
well happen that some of the intersections of Eq.~(\ref{product}) are
empty. This means that the transitions between certain pairs of basic
regions are prohibited.  The information about allowed and prohibited
sequences is contained in a transition matrix
\begin{equation}
t_{ij}=\left\{
\begin{array}{cc} 
 1 & \mbox{if} \; f(R_i) \cap R_j \ne \emptyset \\
 0 & \mbox{otherwise} 
\end{array} \right. \; .
\end{equation}
In this way one has set up a one-to-one correspondence between phase
space points and allowed sequences. (The case of different sequences
being associated to the same point is taken care of by identifying such
sequences, and working in a quotient space.) The 
matrix $t_{ij}$ establishes the grammar rules that forbid certain 
sequences of symbols. When $t_{ij}$ is of finite size the dynamics
becomes topologically conjugate to a subshift of finite type.

The existence of a symbolic dynamics allows for an exhaustive coding
of the orbits of the map. In particular, periodic orbits are in
correspondence with the periodic sequences of the same periodicity.
Given an arbitrary system, it is a hard task to decide
if it admits a symbolic dynamics; even if it does, the translation
from symbols to phase space coordinates is in general extremely
difficult. 
The example we will consider (the cat map) does not present
any of these difficulties, thus eliminating non-essential
complications.

In the following we show how the symbolic dynamics of a classical map
can be used to decompose the traces of the quantized map.  The quantum
analogues of the elements of the classical generating partition are the
quantum rectangles $\hat R$ described in Sections
\ref{section2} and \ref{section3}.  The quantum partitions are obtained by
translating to quantum mechanics the steps in the construction of the
classical ones. Starting from the quantizations of the regions of the
classical basic partition, we define the quantum refinement in two
steps.  First the regions (quantum ``projectors") are propagated using
the Heisenberg equations of motion. Then, noting that ``intersections"
of quantum rectangles correspond to matrix multiplications, we arrive
at a quantum product partition with elements written as a time ordered
multiplication of matrices
\begin{eqnarray}
\label{markquan}
\hat R_{\nu(-K,M)} 
     & = & 
\hat U^{-K} \hat R_{\epsilon_{-K}} \hat U^{K} 
\ldots 
\hat U^{M} \hat R_{\epsilon_M} \hat U^{-M}       \nonumber \\  
     & = &
\hat U^{-K} \hat R_{\epsilon_{-K}} \hat U 
           \hat R_{\epsilon_{-K+1}}  \ldots  
       \hat R_{\epsilon_{M-1}} 
\hat U \hat R_{\epsilon_M} \hat U^{-M} \; .
\end{eqnarray}
The counterpart of the classical decomposition of the phase space is
the quantum decomposition of the identity
\begin{equation}
\label{normalization}
\sum_{\nu(-K,M)} \hat R_{\nu(-K,M) } = \openone/N \; ,
\end{equation}
$N$ being the dimension of the Hilbert space. 
The quantum propagation is also a {\em shift\/}:
\begin{equation}
\label{quanshift}
\hat U^{-1} \hat R_{\epsilon_{-K}\ldots \epsilon_{-1} 
\cdot 
\epsilon_{ 0}\epsilon_{1}\ldots \epsilon_{ M} } \hat U=
\hat R_{\epsilon_{-K}\ldots \epsilon_{ 0}
\cdot 
\epsilon_{1}\ldots \epsilon_{ M} } \; .
\end{equation}
Even though the quantum rectangles don't have zero ``intersection", 
in the semiclassical limit, the product of two elements of the partition
tends to the null operator, except from possible singularities due
to border effects. Last, when $N \to \infty$ and $K,M$ fixed, the
quantum rectangles tend to the classical ones. The precise meaning
of this limit, and the way it is achieved were discussed in 
Sec.~\ref{section3}.  

The key property of the quantum partition we have constructed is the
symbolic decomposition of the traces of the propagator. Consider the
discrete path sum for the trace of a power of the propagator 
in the coherent state representation
\begin{equation}
\label{pathcoh}
\mbox{Tr} \hat U^{L}=\frac{1}{N^{2L}} 
\sum 
\langle \alpha_0 | \hat U | \alpha_{L-1} \rangle \langle \alpha_{L-1} |
                   \hat U \ldots
                          | \alpha_{  1} \rangle \langle \alpha_{  1} |
       \hat U | \alpha_0 \rangle \; 
\end{equation}
where the sum runs over all the {\em closed} paths
$\alpha_0,\alpha_1,\ldots,\alpha_{L-1},\alpha_{L} \equiv \alpha_{0}$,
which are discrete both in time and in the coordinates (we recall that
$\alpha \equiv q+ip$ moves on the discrete $q$-$p$ grid).
Semiclassically the trace of $\hat U^{L}$ will be dominated by the
periodic trajectories (of period $L$) of the classical map $U$ and
their neighboring paths. Symbolic dynamics allows for classifying not
only the trajectories but also the paths according to their symbolic
history. So, one has a natural way of partitioning the space of paths
into disjoints subsets, each one characterized by a symbol $\nu$ of
length $L$ and containing the periodic trajectory $\ldots \nu \nu
\ldots$. But this mechanism of path grouping is automatically
implemented by the quantum projectors:
\begin{eqnarray}
\label{pathrec}
\mbox{Tr} \hat U^{L}  &   =   & \mbox{Tr}
\sum_{\nu} 
\hat U \hat R_{\epsilon_{L-1}} \ldots
\hat U \hat R_{\epsilon_{  1}}
\hat U \hat R_{\epsilon_{  0}} \nonumber \\\
                     & \equiv & 
\sum_{\nu} \mbox{Tr} \hat U^L_{\nu} \; . 
\end{eqnarray}
The $\hat R$'s are the quantum regions associated to the coherent
representation and now the sum runs over the sequence labels
$\nu=\epsilon_0 \epsilon_1 \ldots \epsilon_{L-1}$.
Eq.~(\ref{pathrec}) is completely equivalent to (\ref{pathcoh}),
the difference being just the grouping of closed paths into families 
sharing the same symbolic code $\nu$. Each one of these families 
contributes to a partial trace $\mbox{Tr} \hat U_{\nu}$.
Analogous results are obtained in
the Kirkwood case. In fact, starting from a path sum in the Kirkwood
representation,
\begin{equation}
\label{pathkirk}
\mbox{Tr} \hat U^{L}=\frac{1}{N^{L}} 
\sum 
\langle q_0 | p_0 \rangle \langle p_0 | \hat U \ldots \hat U
| q_1 \rangle \langle q_1 |
 p_1 \rangle \langle p_1 | \hat U | q_0 \rangle  \; , 
\end{equation}
one arrives at the same result of Eq.~(\ref{pathrec}) but with the
Kirkwood rectangles instead of the coherent ones.  
Using the cyclic property, the partial traces of 
(\ref{pathrec}) (or the Kirkwood counterparts)
can be rewritten in terms of the refined rectangles of
Eq.~(\ref{markquan})
\begin{equation}
\label{partial}
\mbox{Tr} \, \hat U^L_{\nu(K,M)} =
\mbox{Tr} \left[   \hat U^L \hat R_{\nu(K,M)} \right] \; .
\end{equation}
The integers $K,M$ must satisfy $K+M=L-1$, but are otherwise
arbitrary.  By varying $K$ and $M$ ($L$ fixed) one constructs different
types of rectangles, e.g., the choice $K$=0, $M$=$L$-1 produces
``unstable" rectangles (stretched along the unstable manifolds)
\begin{equation}
\hat R_{ \cdot 
\epsilon_{ 0} \epsilon_{1}\ldots \epsilon_{L-1} } = 
               \hat R_{\epsilon_{ 0}} 
   \hat U^{ 1} \hat R_{\epsilon_{ 1}} \hat U^{-1}
               \ldots 
  \hat U^{L-1}  \hat R_{\epsilon_L} \hat U^{-(L-1)} \; . 
\end{equation}
Similarly, with $M=0$ and $K$=$L$-1, ``stable" rectangles are obtained.
Anyway, stable and unstable rectangles are related by the unitary 
transformation (\ref{quanshift}), ensuring that 
$\mbox{Tr} \hat U^L_{\nu(K,M)}$ does not depend on the
particular choice of $K,M$. Moreover, the trace of each symbolic piece
is cyclically invariant [as is obvious from (\ref{pathrec})] and 
therefore the
decompositions into invariant cycles in one-to-one correspondence
with periodic orbits of the map.

The refined rectangle $\hat R_{\nu(K,M)}$ has as classical limit the
characteristic function of the classical region $R_{\nu(K,M)}$.
Thus, its role in (\ref{partial}) consists essentially in cutting the
matrix $\hat U^L$ into pieces. The Kirkwood rectangles act onto the 
Kirkwood matrix $\langle p |\hat  U^L | q \rangle$:
\begin{eqnarray}
\mbox{Tr} \left (\hat U^L \hat R_{K,\nu} \right)  & = & 
                   \sum_{q,p} \langle p | \hat U^L | q \rangle 
                              \langle q | \hat R_{K,\nu} | p \rangle 
 \nonumber \\
                              & \approx &
               \frac{1}{N} \sum_{q,p} \langle p | \hat U^L | q \rangle 
                              \Delta_{R_\nu}(q,p)            \; . 
\end{eqnarray}
The coherent rectangles perform a similar action but on the operator 
symbol
$\langle \alpha | \hat U^L | \alpha \rangle$: 
\begin{eqnarray}
\mbox{Tr} \left( \hat U^L \hat R_{z,\nu} \right) & \approx & 
                   \mbox{Tr} \left(\hat U^L \sum_{\alpha \in R_\nu} 
                              | \alpha \rangle \langle \alpha | \right)
 \nonumber \\
                              & = &
         \frac{1}{N^2}          \sum_{\alpha}
                   \langle \alpha | \hat U^L | \alpha \rangle 
                   \Delta_{R_\nu}(\alpha)     \; . 
\end{eqnarray}
In both cases the semiclassical partial trace is obtained by summing
over that piece of the matrix which corresponds to the classical
rectangle. Thus each symbolic piece captures the local structure
of the propagator in the vicinity of a periodic point labeled
bu $\nu$ and by stationary phase yields the Gutzwiller-Tabor
contribution of the corresponding periodic orbit. Forbidden symbols
lead to semiclassically small contributions\cite{gutzwiller90}.

The symbolic decomposition we have presented has the nice feature of
reducing the problem of understanding the asymptotic limit of the
traces of the propagator to the analysis of individual ``partial"
traces $\mbox{Tr} \hat U^L_{\nu}$, 
each one characterized by a code given by the symbolic
dynamics, and ruled by a periodic point.

\subsection{A numerical application}

The simplest system in which the quantum partitions can be applied
to decompose the propagators is perhaps the baker's
map \cite{saraceno94}. Its generating partition consists in two 
rectangles, which, due to the fact that the expanding and contracting
directions are parallel to the coordinate axes,
are solely defined by conditions on $q$. 
As a consequence, the quantum rectangles for the baker's reduce 
{\em exactly} to projectors on subspaces \cite{saraceno94}. This greatly
simplifies the symbolic analysis of the quantum baker's, allowing
very detailed studies of its partial traces
\cite{saraceno94,toscano97}.

However, the baker's is too special for illustrating the properties
of the rectangles: many of them are satisfied trivially. 
Moreover,
the partial traces of the baker's display some unpleasant anomalies
that difficult the semiclassical analyses \cite{saraceno94,toscano97}.

\begin{figure}[ht]
\hspace{-1.0pc}
\vspace{ 1.0pc}
\epsfxsize=8.5cm
\epsfbox[80 253 532 622]{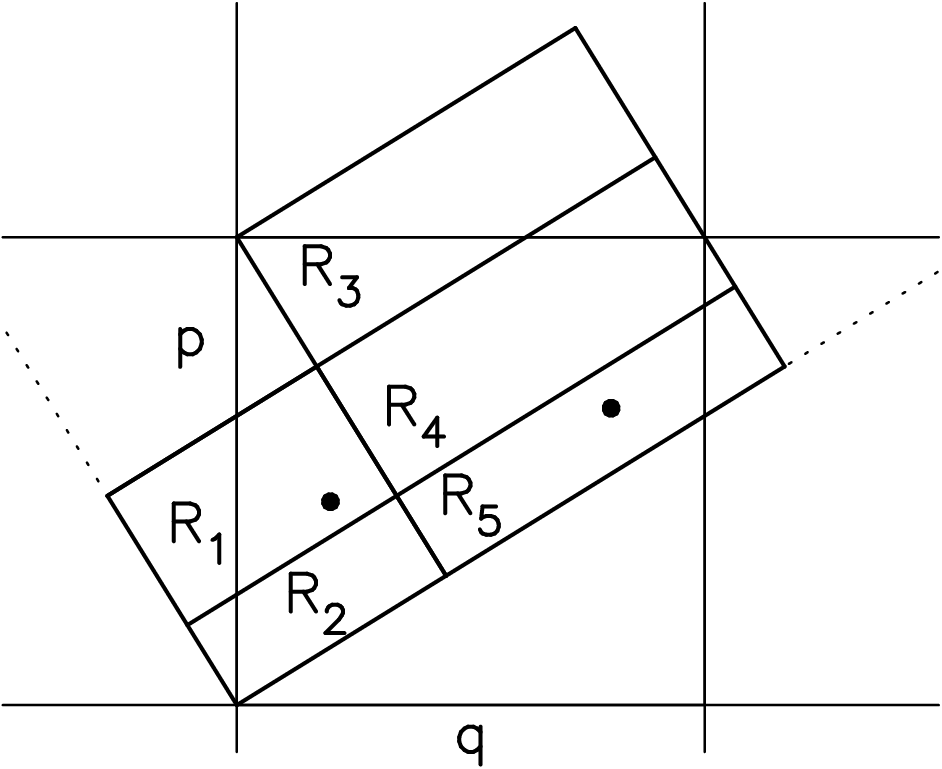}
\caption{The five rectangles of the generating partition of Arnold's cat 
map. Shown are also (parts of) the invariant manifolds of the fixed point 
in the origin (dashed lines), and the period two trajectory with symbolic
label $\nu=51$. }
\label{fig6}
\end{figure}

Still simple enough, the Arnold's cat map $U$ \cite{arnold89} 
is more appropriate for a general illustration of the method and
can be investigated numerically. The classical cat map is defined by
\begin{equation}
\label{catclas}
\left(
\begin{array}{c}
 q' \\ p'
\end{array}
\right)=
\left(
\begin{array}{cc}
 2 & 1 \\
 1 & 1
\end{array}
\right) 
\left(
\begin{array}{c}
 q \\ p
\end{array}
\right) ~ \mbox{mod 1} \; .
\end{equation}
This is a linear, hyperbolic, and continuous map of the torus. 
As its invariant manifolds are not aligned with the coordinate axes, 
the rectangles of the generating partition \cite{adler70}
(shown in Fig.~\ref{fig6}) are not projectors.
This makes the cat map non-trivial for our purposes.

Before proceeding, we must point out that 
the quantum cat map presents one very particular feature: Gutwiller's 
semiclassical formula gives the exact traces \cite{keating91a}. 
For this reason the cat map is not suitable for studying corrections to
the trace formula. In principle, any decomposition into partial traces
will introduce errors which, however, will cancel out when added up to
produce the whole trace. Thus this model may be useful as a test of the
mechanisms that lead to such cancellation.

Let's now go to the details of the numerical example. The generating
partition of the cat map consists of the five rectangles of
Fig.~\ref{fig6}, which, together with the
``grammar rules" embodied in the transition matrix
\begin{equation}
t_{ij} = 
\left(
\begin{array}{ccccc}
 0 1 0 0 1 \\
 0 1 0 0 1 \\
 1 0 1 1 0 \\
 1 0 1 1 0 \\
 1 0 1 1 0 
\end{array}
\right)
\end{equation}
define the symbolic dynamics of the cat\cite{adler70}.
For simplicity we will restrict our analysis to the decomposition of
the trace of the time-two propagator
\begin{equation}
\mbox{Tr} \hat U^2=
\sum_{\epsilon_0,\epsilon_1=1}^5 \mbox{Tr} 
\left(
\hat U \hat{R}_{\epsilon_0} \hat U \hat{R}_{\epsilon_1} 
\right)  \; . 
\end{equation}
The rectangles $\hat R_\epsilon$ are the quantum versions of the
regions of Fig.~\ref{fig6} and can be constructed from either the
coherent state representation or Kirkwood's. The construction of the
quantum propagator $\hat U$ for linear automorphisms of the torus is
presented in \cite{berry80} [notice that Arnold's cat
(\ref{catclas}) is only quantizable for $N$ even].

Each partial trace can be written asymptotically as a
Gutzwiller term plus corrections that go to zero as $N \to \infty$:
\begin{equation}
\label{traceformula}
\mbox{Tr} \hat U^2_{\nu} = 
A_{\nu} \exp \left( 2 \pi i N S_\nu \right) + 
\delta_\nu(N)  \; ,
\end{equation}
where $A_\nu$ is the amplitude and $S_\nu$ the action of the periodic
orbit \cite{tabor88}. We remark that in the case of cat maps the
corrections $\delta_\nu$ will cancel out exactly when summing over $\nu$ 
because the semiclassical 
trace formula is exact in this special case. 
In general this will not be true, and the method allows to study the
corrections coming from each periodic orbit.

In order to quantify the errors associated to the symbolic partition of
the space of paths, we study numerically the semiclassical limit of one 
element of the partition, namely $\mbox{Tr}U^2_{51}$. This trace is 
dominated by the periodic trajectory shown in Fig.~\ref{fig6} and its
neighborhood; its asymptotic limit is the Gutzwiller formula
(\ref{traceformula}) with $A_{51}=1/\sqrt{5}$ and $S_{51}=3/10$
\cite{keating91b}.

We can understand the asymptotic behavior of the corrections 
$\delta_\nu$
by recalling that our decomposition essentially amounts to cutting 
the matrix of $\hat U$ into rectangular blocks. 
Let's first estimate the corrections in the Kirkwood's case. 
The Kirkwood matrix of $U^L$ has constant amplitude \cite{berry80}
and phase that 
oscillates rapidly except in the vicinity of the fixed points of $U^L$
\cite{saraceno94}. 
Computing the partial trace amounts to summing up the 
matrix elements $\langle p | U^2 | q \rangle $ that
lie inside the region $R_{51} \equiv R_5 \cap U^{-1}(R_1)$ 
(shown in Fig.~\ref{fig5}).
In the semiclassical
limit we can replace the sum by an integration and do the latter
using the stationary phase method. In this approximation we must only
take into account the contributions of the {\em critical points}
\cite{mandel95}.
The most important contribution comes from the  
the periodic orbit (critical point of the first kind) and its
neighborhood. This gives rise to the Guzwiller term, which is of
order zero in $\hbar$ [${\cal O}(\hbar^0)$]. 
The corrections $\delta_{51}(N)$ are associated to critical points of 
second and third
kind. The critical points of the 
second kind, i.e., points where the phase is stationary with respect
to displacements along the borders of the rectangle, 
contribute with terms  ${\cal O}(\hbar^{1/2})$. The corners 
(third kind critical points) contribute with terms 
${\cal O}(\hbar^{3/2})$. (In the baker's the situation is more
complicated because of the coalescence of critical points of different
kind, namely some fixed points lie on the borders
of the rectangles. These anomalous points
give rise to terms ${\cal O}(\log \hbar)$ \cite{saraceno94,toscano97}.) 
Having exhausted the critical points, we 
conclude that the border errors in Kirkwood's representation 
are ${\cal O}(\hbar^{1/2})$.
On the other hand, in the coherent case, one expects the amplitudes
$\langle \alpha | \hat U^L | \alpha \rangle$ to decay exponentially fast
as one moves away from the classical trajectory. 
The phases do still oscillate fast. 
However, due to the exponential damping, the border effects in the 
coherent decomposition should then be 
${\cal O} [ \hbar^{1/2} \exp(-C^2/\hbar)]$, where $C$ is proportional
to the distance from the fixed point to the border. 
Of course, this regime will only
be reached once the stationary phase neighborhood of the fixed point
[whose radius is ${\cal O} (\hbar^{1/2})$] is completely contained in 
$R_{51}$.

For the coherent case we calculated numerically the correction 
$\delta_{51}$ as a function of $N$. Up to $N=100$ we computed the 
partial trace exactly, i.e., 
\begin{equation}
\frac{1}{N^4}
\sum_{\alpha \in R_1, \, \beta \in R_5} 
\langle \alpha | \hat U | \beta  \rangle 
\langle \beta  | \hat U | \alpha \rangle   \; .
\end{equation}
From then on, due to computer time limitations, we resorted to a local
semiclassical approximation for the coherent-state propagator.
This is equivalent to replacing the torus propagator
$\langle \alpha | \hat U^2 | \beta \rangle$ by a 
plane propagator which is the quantization of the linear
dynamics in the vicinity of the period-two trajectory $\nu=51$. 
The errors introduced in this approximation arise from ignoring the
contributions of ``sources" located at equivalent (mod 1)
positions in the plane \cite{berry80}. These errors are also
${\cal O} [ \hbar^{1/2} \exp(-C'^2/\hbar)]$, but with $C'$ 
much larger than $C$, and thus can be neglected. Once the partial
trace was calculated, we obtained the correction $\delta_\nu$ by 
subtracting the Gutzwiller term.

\begin{figure}[ht]
\hspace{-1.0pc}
\vspace{ 1.0pc}
\epsfxsize=8.5cm
\epsfbox[45 165 476 549]{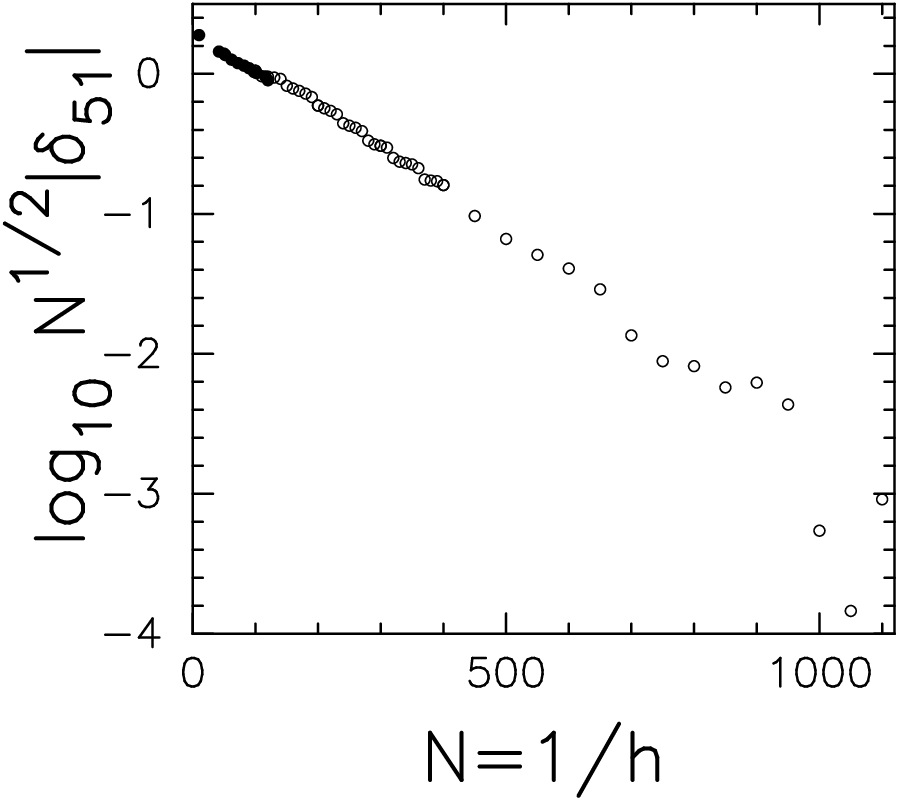}
\caption{Corrections to the Gutwiller trace formula, $\nu=51$. }
\label{fig7}
\end{figure}

In Fig.~\ref{fig7} we show the numerical results in a way that
permits a direct comparison with our analytical considerations
above. In fact, the log-linear plot suggests that the corrections
$\delta_\nu$ in the coherent state decomposition are indeed
exponentially small in the semiclassical parameter $1/\hbar$.
Accordingly, the decomposition which uses rectangles constructed
from the Kirkwood representation introduces border errors of order
$\hbar^{1/2}$. 
We recall that Gutzwiller's trace formula is exact for the cat maps.
For typical maps one expects corrections to this formula of order
$\hbar^k$, with $k \ge 1$; e.g., $k=1$ for the {\em perturbed} cat maps
\cite{boasman94}.  Both Markov partitions considered here, either based
on coherent-state or Kirkwood rectangles, allow to study such
corrections term by term.  In the coherent case, the partitioning of
the space of paths does not introduce significant border effects, given
that the contributions of neighboring paths decrease exponentially as
one moves away from the central trajectory. On the other side, the use
of a sharp representation like Kirkwood's produces non-negligible
boundary contributions to each partial trace. Of course these boundary
terms will cancel out when the partial traces are summed up to give the
whole trace.  Even though, they have to be carefully identified to
isolate the genuine partial corrections to the Gutzwiller trace
formula.

\section{Concluding remarks}

We have begun the application of symbolic dynamics techniques, essential in
classical chaotic problems, in quantum mechanics.  As a fist step we
constructed quantum analogues to regions of classical phase space: they
are the quantizations of the characteristic functions of the classical
regions. We have used Kirkwood's and a coherent state representation.
The study of metrical and spectral properties show that they behave
asymptotically as projectors over those regions. They also present the
diffraction effects typical of ondulatory phenomena.

For a finite-type subshift, the quantization of the rectangles of the
classical generating partition gives rise to a quantum partition which
induces a symbolic decomposition of the propagator. This partition
allows for writing a trace formula which is both exact and structurally
identical to the Gutzwiller trace formula. Thus the problem  of
understanding the semiclassical limit of the traces of a propagator is
reduced to the analysis of partial traces coded by the symbolic
dynamics.  The objects we have constructed tend asymptotically to their
classical counterparts and respond to same dynamics. In this way, one
can verify step by step many manipulations that up to now could only be
done at a semiclassical level.


Before concluding we would like to emphasize that the construction
presented here is by no means restricted to phase space regions that
are Markov partitions.  Any region of phase space selected for
"attention" can be handled in the same way and its quantum properties
explored. For example, if a closed problem is turned into a scattering
one by the removal of a section of the boundary or the attachment of a
soft wave guide the decomposition leads to the consideration of coupled
interior and closure problems projected from the corresponding phase
space regions \cite{ozorio98}. Another application is to think of the
phase space projectors as ``measurements'' occurring along the quantum
history of the system, and the associated decoherence that result.


\acknowledgements

The authors have benefited from discussions with 
E. Vergini, A. Voros, and A. M. Ozorio de Almeida.
R.O.V acknowledges Brazilian agencies FAPERJ and PRONEX for financial
support, and the kind hospitality received at the Centro
Brasileiro de Pesquisas F\'{\i}sicas and at Laboratorio TANDAR, where
part of this work was done. Partial support for this project was
obtained from ANPCYT PICT97-01015 and CONICET PIP98-420.



\end{multicols}

\end{document}